\documentclass[11pt, letterpaper]{article}

\usepackage[table]{xcolor}
\usepackage[margin=1in]{geometry}
\usepackage{algorithmic, algorithm}
\usepackage[english]{babel}
\usepackage{amssymb}
\usepackage{amsmath}
\usepackage{xcolor}
\usepackage{cite}
\usepackage{graphicx}
\usepackage{amsthm}
\usepackage{dirtytalk}
\usepackage[hidelinks]{hyperref}
\usepackage{tikz}
\usetikzlibrary{positioning}

\theoremstyle{definition}
\newtheorem{theorem}{Theorem}
\newtheorem{lemma}{Lemma}
\newtheorem{definition}{Definition}
\newtheorem{corollary}{Corollary}

\newtheorem{construction}{Construction}

\newcommand{\set}[2]{\left\{#1\;\left|\; #2\right.\right\}}
\newcommand{\abs}[1]{\left|#1\right|}
\renewcommand{\v}[1]{\mathbf{#1}}

\title{Universal Framework for Parametric Constrained Coding}

\author{Daniella Bar-Lev, Adir Kobovich, Orian Leitersdorf, and Eitan Yaakobi}
\date{}

\begin{document}

\maketitle

\begin{abstract}
Constrained coding is a fundamental field in coding theory that tackles efficient communication through constrained channels. While channels with fixed constraints (e.g., a fixed set of substrings may not appear in transmitted messages) have a general optimal solution, there is increasing demand for supporting \emph{parametric} constraints that are dependent on the message length and portray some property (e.g., no $\log(n)$ consecutive zeros). Several works have tackled such parametric constraints through \emph{iterative} algorithms, yet they require complex constructions specific to each constraint to guarantee convergence through \emph{monotonic progression}. In this paper, we propose a universal framework for tackling \emph{any} parametric constrained-channel problem through a novel \emph{simple} iterative algorithm. By reducing an execution of this iterative algorithm to an acyclic graph traversal, we prove a surprising result that guarantees convergence with efficient average time complexity \emph{even without requiring any monotonic progression}.

We demonstrate the effectiveness of this universal framework by applying it to a variety of both \emph{local} and \emph{global} channel constraints. We begin by exploring the local constraints involving illegal substrings of variable length, where the universal construction essentially iteratively replaces forbidden windows. We apply this local algorithm to the minimal periodicity, minimal Hamming weight, local almost-balanced Hamming weight and the previously-unsolved minimal palindrome constraints, and demonstrate state-of-the-art results through simple adaptations of the universal algorithm. We then continue by exploring global constraints, and demonstrate the effectiveness of the proposed construction on the repeat-free encoding, reverse-complement encoding, and the open problem of global almost-balanced encoding. For reverse-complement, we also tackle a previously-unsolved version of the constraint that addresses overlapping windows. Overall, the proposed framework generates state-of-the-art constructions with significant ease while also enabling the simultaneous integration of multiple constraints for the first time. 
\end{abstract}

\section{Introduction}
\label{sec:introduction}

Constrained coding is a fundamental field with a wide range of applications including magnetic and optical storage, flash memory, DNA storage, wireless communication, and satellite communication~\cite{immink2004codes, marcus2001introduction, immink2022innovation}. These applications are subject to unique challenges and constraints that must be satisfied by the data encoding. For instance, magnetic storage constrains a bounded number of zeros between consecutive ones due to voltage interference in the sensing circuitry that may otherwise lead to errors~\cite{immink1990runlength}. In flash memory, constrained codes are used in order to mitigate the inter-cell interference between the nearby wordlines and bitlines~\cite{qin2014constrained, bermanflash}. The recently-emerging field of DNA-based data storage~\cite{erlich2017dna} (storing digital information as synthetic DNA strands) similarly enforces constraints on the encoded data to increase molecular stability and reduce errors in synthesis and sequencing~\cite{bornholt2016dna, erlich2017dna, heckel2019characterization, organick2018random}. Overall, the field of constrained coding deals with the development of efficient constructions that encode general information into messages that satisfy the channel constraints.

Constrained coding has its roots in solving \emph{fixed constraints} -- constraints that are independent of the message length. For example, \emph{local} fixed constraints involve a fixed set of strings that cannot appear as substrings in the encoded data (e.g., the sequence \say{$0011$} cannot appear anywhere in the encoded message). Such constraints have been thoroughly studied in the past, and there exists a universal framework for developing a construction for any such fixed local constraint~\cite{marcus2001introduction}. The process begins by constructing a deterministic finite automaton that accepts all the constrained words, and then utilizing the state-splitting algorithm~\cite{adler1983algorithms} to design encoders with rate approaching the capacity. Further, the channel capacity may be computed using Perron–Frobenius theorem~\cite{mincnonnegative} that relates the capacity to the spectral radius of the automaton adjacency matrix.

Conversely, there is an increasing demand for \emph{parametric} constraints that portray some general property and are dependent on the message length. For example, a \emph{local} parametric constraint on a message of length $n$ may be that all substrings of length $\ell(n)$ possess Hamming weight of at least $p(n)$. This form of constraints currently lacks a unified universal framework, and instead existing methods are often tailored to specific applications -- requiring significant effort when faced with new constraints or when attempting to enforce several constraints simultaneously. Specifically in the case of \emph{local} parametric constraints which enforce restrictions on $\ell(n)$-substrings of the overall encoded message, previous works~\cite{van1997construction, van1998frame, van2010construction, nguyen2020binary, schoeny2017codes, immink2017design, immink2020properties, levy2018mutually} have typically followed a concept known as \say{sequence replacement}~\cite{van1997construction}. Essentially, forbidden substrings are \emph{iteratively} removed and replaced with an alternative encoding at the end of the string consisting of a shortened encoding of the forbidden substring and the index of the substring in the encoded message. Further, a variation of sequence replacement has been used for global constraints involving pairs of substrings~\cite{elishco2021repeat, nguyen2023design}. Yet, the difficulty arises from proving the convergence of the iterative encoder due to the inter-dependence between the substrings: replacing one forbidden substring may cause a different substring to become forbidden. Therefore, complex modifications are tailored for each constraint to guarantee \emph{monotonic progression} by exploiting properties of the constraint; for example, one approach is to cause the message to \say{shrink} during the iterative steps and then to extend the message with a carefully-chosen suffix that both satisfies the constraint and can be identified by the decoder~\cite{van2010construction, nguyen2020binary, elishco2021repeat, nguyen2023design}. Thus, there is a lack of a unifying framework for parametric constraints overall, and even in the special case of \emph{local} parametric constraints we require complex encodings specific to each channel constraint. 

In this work, we propose the first \emph{universal framework} for parametric constrained coding problems and demonstrate its effectiveness on a variety of local and global channel constraints. At the core of the framework is a very \emph{simple} algorithm inspired by a recent work on constrained periodicity~\cite{ConstrainedPeriodicity} that does \emph{not} possess monotonic progression and yet nonetheless converges efficiently due to a proof involving a reduction to an acyclic graph traversal. \emph{Due to the simplicity of the algorithm, it is highly general and applicable to a variety of channels.} We begin by presenting the core algorithm in the most general form, and then continue by presenting two useful general cases of one-symbol-redundant constructions and the \emph{first} general intersection method for designing constructions that simultaneously satisfy multiple constraints. We apply this framework to a practical variety of both local and global constraints, demonstrating improved parameters compared to previous works~\cite{van1997construction, van1998frame, van2010construction, nguyen2020binary, schoeny2017codes, immink2017design, immink2020properties, levy2018mutually} with algorithms that are drastically simpler since the convergence is automatically guaranteed by the framework. Further, we also tackle the previously-unsolved local constraint of minimal palindromes, and utilize it as part of the intersection method to propose a hybrid local-global construction for a previously-unsolved constraint in DNA storage~\cite{warris2018correcting, nguyen2023design}. Lastly, we tackle the open problem of the global almost-balanced Hamming weight constraint through a combination of the universal framework and the intersection method with a technique inspired by arithmetic coding~\cite{rissanen1979arithmetic}. 

Overall, this paper contributes the following:
\newpage
\begin{enumerate}
    \item \textbf{Universal Framework:} In Construction~\ref{const:universal}, we propose the first universal framework that may be readily applied to any parametric constrained coding problem -- guaranteeing efficient convergence even without monotonic progression. Specifically, given an \emph{injective} \emph{step} function that merely maps a message that does not satisfy the constraint to any non-starting message, the framework constructs a construction for the constraint that converges with efficient average time complexity. We prove this convergence for any channel constraint by demonstrating that the encoder's iterations are analogous to a graph traversal, and then demonstrating that there are no cycles reachable in the graph from the starting nodes. 
    \item \textbf{One-Symbol-Redundant Framework:} A popular form of parametric constrained codes involves constructions that require only a single symbol of redundancy. In Construction~\ref{const:shrink}, we present a simplified version of Construction~\ref{const:universal} for a single redundancy symbol that is even simpler to adopt. Overall, we find that the construction receives an enumeration of the forbidden messages and returns an enumeration on the legal messages -- we show in the later examples that it is typically far simpler to find enumerations on the forbidden messages.
    \item \textbf{Intersection Method:} The \emph{simultaneous} adherence to multiple different parametric channel constraints is of particular interest to applications such as DNA storage~\cite{erlich2017dna}. Previous works have struggled with developing constructions that satisfy multiple constraints simultaneously as the independent constructions for each constraint vary due to the customization previously required for each constraint. Yet, in this paper we provide a general method in Construction~\ref{const:intersect} that automatically enables the integration of multiple constraints (each implemented in the universal framework) into a single construction -- this is possible since the independent constructions are all based on the same universal framework.
    \item \textbf{Local Constraints:} We apply the universal framework to \emph{local} constraints that operate on substrings of the transmitted message, thereby providing a general construction for local constraints. We then utilize this construction on a variety of practical local channels (minimal Hamming weight, local almost-balanced Hamming weight, and minimal periodicity) and demonstrate improved parameters over previous state-of-the-art constructions with far simpler algorithms. We also tackle the previously-unsolved minimal palindrome constraint, which we then later utilize to improve a different global constraint. 
    \item \textbf{Global Constraints:} We demonstrate the effectiveness of the universal framework on the more challenging \emph{global} constraints that are defined on the entire message. We begin with the repeat-free constraint, where we improve the state-of-the-art results for both overlapping and non-overlapping repetitions. We then continue with the secondary structure (DNA reverse complement constraint), where previous works only successfully tackled non-overlapping substrings and we demonstrate significant improvement for that case. Further, by utilizing the intersection method, we are able to tackle overlapping windows as well for the first time with similar parameters. We conclude by tackling the open problem of the almost-balanced weight constraint~\cite{nguyen2020binary}, presenting an efficient algorithm with one redundancy symbol. 
\end{enumerate}

The paper is organized as follows. Section~\ref{sec:universal} provides preliminary definitions and then proposes and proves the universal framework for parametric channel constraints along with the general extensions. Section~\ref{sec:local} applies this framework to local parametric constraints, and Section~\ref{sec:global} tackles global parametric constraints. Lastly, Section~\ref{sec:conclusion} concludes this paper.

\section{Universal Framework}
\label{sec:universal}

This section proposes a new universal framework for \emph{parametric} constrained coding that consists of non-monotonic iterative constructions which are nonetheless guaranteed efficient average-time complexity. We begin with several preliminary definitions for parametric constrained channels, and then continue by proposing the fundamental iterative construction proposed in this paper. We then prove the convergence of this algorithm through a reduction to an acyclic graph traversal, and demonstrate efficient average time complexity. We conclude with two interesting general special cases of this construction, where the next sections utilizing them towards concrete applications. 

\subsection{Parametric Channel Constraints}
\label{sec:universal:channel}

We explicitly define in this section channel constraints that are of parametric form. Parametric constraints differ from fixed constraints in that they are described according to a general formulation that generalizes to messages of different lengths. For example, a parametric channel constraint may be that messages of length $n$ may not contain $\log (n)$ consecutive zeros. We generalize this as follows,

\begin{definition}[Parametric Channel Constraint]
A parametric channel constraint $\mathcal{C}$ applied to a channel of length $n$ is a channel that may only transmit messages $\v{y} \in \Sigma^n$ such that $\v{y} \in \mathcal{C}(n)$.
\end{definition}

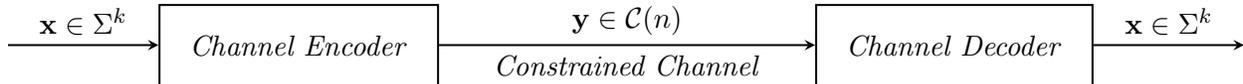
\begin{figure}[t]
    \centering
    \begin{tikzpicture}
 
    \node [draw,
        inner sep=0.4cm,
        line width=0.75pt,
    ]  (encoder) {\emph{Channel Encoder}};
    
    \node [draw,
        inner sep=0.4cm,
        right=5cm of encoder,
        line width=0.75pt,
    ]  (decoder) {\emph{Channel Decoder}};

    \draw[-stealth, line width=0.75pt] ([xshift=-2cm]encoder.west) -- (encoder.west)
        node[midway, above]{$\v{x} \in \Sigma^k$};
    
    \draw[-stealth, line width=0.75pt] (encoder.east) -- (decoder.west) 
        node[midway, above]{$\v{y} \in \mathcal{C}(n)$}
        node[midway, below]{\emph{Constrained Channel}};
        
    \draw[-stealth, line width=0.75pt] (decoder.east) -- ([xshift=2cm]decoder.east)
        node[midway, above]{$\v{x} \in \Sigma^k$};
     
    \end{tikzpicture}
    \caption{Construction (encoder and decoder) for parametric channel constraint $\mathcal{C}$ applied to a channel of length $n=k+r$. The constrained channel only allows transmission if $\v{y} \in \mathcal{C}(n)$.}
    \label{fig:channel}
\end{figure}

In this work, we focus on constructing efficient constructions (encoders and decoders) for parametric channel constraints. As seen in Figure~\ref{fig:channel}, these constructions encode all messages of length $k$ into messages of length $n=k+r$ that satisfy the parametric channel constraint for channels of length $n$. We explicitly define the channel encoder and channel decoder as follows,

\begin{definition}[Channel Encoder]
A parametric channel encoder $f_k : \Sigma^k \rightarrow \mathcal{C}(n)$ encodes general messages of length $k$ to messages of length $n=k+r$ that satisfy the parametric constraint $\mathcal{C}$. We define the redundancy of the encoder as $r$.
\end{definition}

\begin{definition}[Channel Decoder]
A parametric channel decoder $g_k : \mathcal{C}(n) \rightarrow \Sigma^k$ satisfies $g_k(f_k(\v{x})) = \v{x}$ for any $\v{x} \in \Sigma^k$.
\end{definition}

\subsection{Proposed Construction}
\label{sec:universal:construction}

In this section, we propose the general universal construction for parametric channel constraints. We first discuss the approach and then continue by explicitly defining the encoder and decoder. 

\begin{figure}[t]
    \centering
    \includegraphics[width=0.8\linewidth]{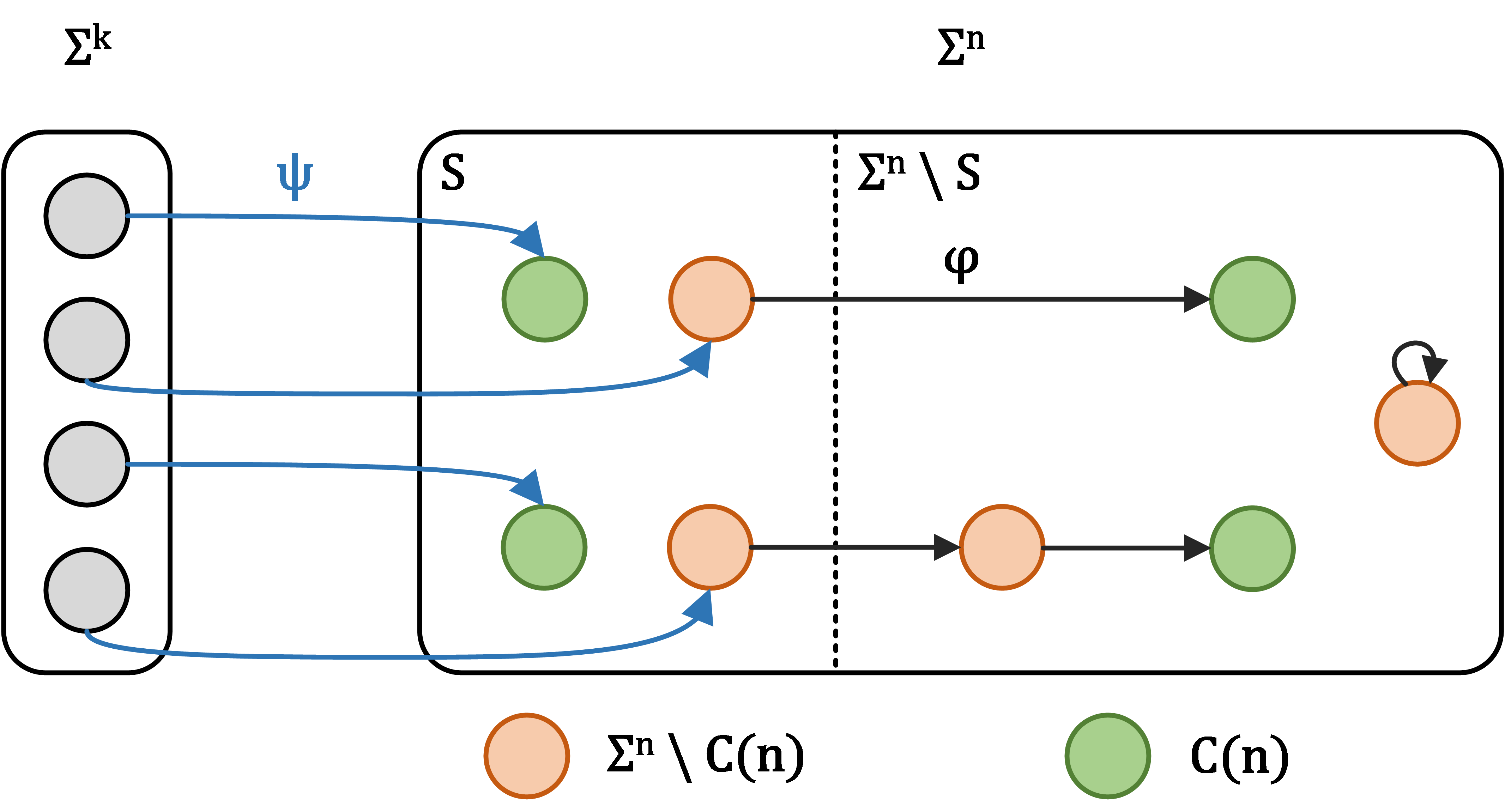}
    \caption{Example configuration of the universal iterative construction for $n=3$ and $k=2$.}
    \label{fig:graph}
\end{figure}

We begin by developing the intuition for the universal iterative construction. Consider $\Sigma^n$, the set of all potential messages of length $n$, as shown in the right of Figure~\ref{fig:graph} (where each node represents a potential message). One possible partition of this space is according to the constraint $\mathcal{C}(n)$, with the nodes in Figure~\ref{fig:graph} colored according to whether they satisfy the constraint. We also consider a secondary partition of the space $S \subseteq \Sigma^n$ of size $\abs{\Sigma}^k$ that will represent the initial states of the algorithm (with $\psi : \Sigma^k \rightarrow S$ some efficient enumeration of $S$). Lastly, we assume the existence of an efficient \emph{injective} step function $\varphi : \overline{\mathcal{C}(n)} \rightarrow \overline{S}$. The encoder operates by first embedding $\v{x} \in \Sigma^k$ in $S$ using $\psi$, and then iteratively applying $\varphi$ to the message until the message belongs to $\mathcal{C}(n)$. The operation of the encoder can be considered as similar to a finite state machine where $S$ represents the initial states and $\mathcal{C}(n)$ represents the accepting states. We notice that the requirements for $\varphi$ are very minimal as we only require it to be an injective function for the overall algorithm to be valid; for example, we notice that $\varphi$ in Figure~\ref{fig:graph} even possesses a self-loop, and yet the algorithm never encounters this state as there is no path from an initial state to the self-loop state. The convergence of the iterative algorithms in previous parametric constrained coding applications~\cite{van1997construction, van1998frame, van2010construction, nguyen2020binary, schoeny2017codes, immink2017design, immink2020properties, levy2018mutually} was always shown through monotonic progression (i.e., the message \emph{approaches} a valid message); yet, here we do not require any such monotonic progression and regardless the overall construction still converges efficiently. 

We formalize the encoder and decoder algorithms in Algorithms~\ref{alg:universalEncoder} and~\ref{alg:universalDecoder}. We begin with the encoder in Algorithm~\ref{alg:universalEncoder}, which utilizes $\psi, \varphi$, and an indicator function $\v{1}_{\mathcal{C}(n)} : \Sigma^n \rightarrow \{0,1\}$ (satisfies $\v{1}_{\mathcal{C}(n)}(\v{y}) = 1$ for $\v{y} \in \Sigma^n$ if and only if $\v{y} \in \mathcal{C}(n)$). The encoder proceeds as described: first $\v{x}$ is embedded in $S \subseteq \Sigma^n$ using $\psi$, and then $\varphi$ is iteratively applied until the message is in $\mathcal{C}(n)$. The decoder in Algorithm~\ref{alg:universalDecoder} reverses the steps of the encoder to derive the original message by utilizing $\varphi^{-1}, \psi^{-1}$ and an indicator $\v{1}_S : \Sigma^n \rightarrow \{0,1\}$. The decoder is essentially \say{unraveling} the steps of the encoder using $\varphi^{-1}$, utilizing the fact that \emph{only} the initial states belong to $S$ to know when to terminate the decoding. Overall, we summarize the construction as follows:

\begin{construction}[Universal]
For a given constraint $\mathcal{C}(n) \subseteq \Sigma^{n}$, and given:
\begin{enumerate}
    \item a subset $S\subseteq \Sigma^{n}$ such that $\abs{\mathcal{C}(n)} \geq \abs{S} \geq \abs{\Sigma}^k$,
    \item an injective embedding $\psi : \Sigma^k \rightarrow S$ and indicator $\v{1}_S : \Sigma^{n} \rightarrow \{0,1\}$,
    \item an injective function $\varphi: \overline{\mathcal{C}(n)} \rightarrow \overline{S}$ and indicator $\v{1}_{\mathcal{C}(n)} : \Sigma^{n} \rightarrow \{0,1\}$,
\end{enumerate}
then Algorithms~\ref{alg:universalEncoder} and~\ref{alg:universalDecoder} construct an efficient construction with $r=n-k$ redundancy symbols and $O(\abs{\Sigma}^r \cdot T(n))$ average time complexity for $T(n)$ the maximal complexity amongst $\varphi, \varphi^{-1}, \v{1}_S, \v{1}_{\mathcal{C}(n)}$.
\label{const:universal}
\end{construction}

\begin{algorithm}[t]
    \centering
    \begin{algorithmic}[1]
    \renewcommand{\algorithmicrequire}{\textbf{Input:}}
    \renewcommand{\algorithmicensure}{\textbf{Output:}}
    \REQUIRE $\v{x} \in \Sigma^k$.
    \ENSURE $\v{y} \in \mathcal{C}(n)$. 
    
    \STATE $\v{y} \gets \psi(\v{x})$.
    
    \WHILE{$\v{1}_{\mathcal{C}(n)}(\v{y}) = 0$}
    
    \STATE $\v{y} \gets \varphi(\v{y})$.
    
    \ENDWHILE
    
    \RETURN $\mathbf{y}$.
    
    \end{algorithmic}
    \caption{Universal Iterative Encoder $f_k$}
    \label{alg:universalEncoder}
\end{algorithm}

\begin{algorithm}[t]
    \centering
    \begin{algorithmic}[1]
    \renewcommand{\algorithmicrequire}{\textbf{Input:}}
    \renewcommand{\algorithmicensure}{\textbf{Output:}}
    \REQUIRE $\v{y} \in \mathcal{C}(n)$ such that $f_k(\v{x}) = \v{y}$ for some $\v{x} \in \Sigma^k$.
    \ENSURE $\v{x} \in \Sigma^k$.
    
    \WHILE{$\v{1}_S(\v{y}) = 0$}
    
    \STATE $\v{y} \gets \varphi^{-1}(\v{y})$.
    
    \ENDWHILE
    
    \RETURN $\psi^{-1}(\v{y})$.
    
    \end{algorithmic}
    \caption{Universal Iterative Decoder $g_k$}
    \label{alg:universalDecoder}
\end{algorithm}

\subsection{Convergence and Time Complexity}
\label{sec:universal:convergence}

We now prove that the proposed encoder and decoder converge with efficient average time complexity by presenting the algorithm execution as analogous to an acyclic graph traversal. The difficulty in proving convergence that previous works~\cite{van1997construction, van1998frame, van2010construction, nguyen2020binary, schoeny2017codes, immink2017design, immink2020properties, levy2018mutually} faced with iterative constructions for constrained channels is that the encoder may be \say{stuck} in an infinite loop if there is no monotonic progression. Therefore, all previous works developed custom complex transition functions for each channel that were specifically targeted at guaranteeing monotonic progression. Conversely, we prove in this section that, even without monotonic progression, Construction~\ref{const:universal} will never reach a cycle and thus will always converge for any $\mathcal{C}(n)$. 

Interestingly, even though there may exist cycles in $\varphi$, the encoder will never reach those cycles. For example, consider the self-loop in Figure~\ref{fig:graph}: a state $\v{y} \in \Sigma^n$ such that $\varphi(\v{y}) = \v{y}$. Theoretically, if the encoder were to reach this $\v{y}$, then the algorithm would never converge as it would be stuck in an infinite loop. Yet, we show below that the encoder always inherently avoids such $\v{y}$ that may lead to infinite loops due to two subtle properties of the construction: the fact that the image of $\varphi$ is contained in $\overline{S}$, and the fact that $\varphi$ is injective. This is utilized in Theorem~\ref{the:universalEncoder} to prove that the encoder will never encounter a cycle of \emph{any} length (not only $\v{y} \in \Sigma^n$ such that $\varphi(\v{y}) = \v{y}$).

\begin{theorem} 
The encoder from Algorithm~\ref{alg:universalEncoder} always converges. 
\begin{proof}

We model the encoder as a traversal on a directed graph $G = (V,E)$ with nodes representing messages and edges representing $\varphi$:
\begin{equation}
V = \Sigma^{n} \quad\quad\quad E = \set{(\v{u}, \v{v})}{\v{u} \notin \mathcal{C}(n),\; \varphi(\v{u}) = \v{v}}.
\end{equation}

The right hand side of Figure~\ref{fig:graph} illustrates an example of this graph. We begin by observing the following properties regarding the in-degree $d_{in}$ of nodes in the graph: $d_{in}(\v{y}) \leq 1$ for any $\v{y} \in \Sigma^n$ since $\varphi$ is injective, and $d_{in}(\v{y}) = 0$ for any $\v{y} \in S$ since the image of $\varphi$ is contained in $\overline{S}$. We now assume by contradiction that the algorithm does not converge for input $\v{x} \in \Sigma^k$. Let $\v{y}^0, \v{y}^1, \hdots$ be the intermediate states encountered by the encoder (where $\v{y}^0 = \psi(\v{x})$), corresponding to the path $\v{y}_0 \rightarrow \v{y}_1 \rightarrow \cdots$ in $G$. Since the state space is finite ($\abs{V} = \abs{\Sigma}^n$) and the traversal does not terminate, then we find that there exist a cycle $\v{y}_i \rightarrow \v{y}_{i+1} \rightarrow \cdots \rightarrow \v{y}_{j-1} \rightarrow \v{y}_j = \v{y}_i$. Since $\v{y}_0 \in S$ implies $d_{in}(\v{y}_0) = 0$, then we find that $\v{y}_0$ is not in the cycle. Thus, the cycle is not reachable from $\v{y}_0$ since the in-degree of all nodes in the cycle is at most one and $\v{y}_0$ is not part of the cycle. This is a contradiction as the cycle is reachable from $\v{y}_0$ using the path $\v{y}_0 \rightarrow \v{y}_1 \rightarrow \cdots \rightarrow \v{y}_i$.
\end{proof}
\label{the:universalEncoder}
\end{theorem}

\begin{theorem}
The decoder from Algorithm~\ref{alg:universalDecoder} is correct.
\begin{proof}
We prove that $g_k(f_k(\v{x})) = \v{x}$ for any $\v{x} \in \Sigma^k$; let $\v{x} \in \Sigma^k$ be given. Let $t(\v{x})$ be the number of iterations performed in $f_k(\v{x})$ (finite due to Theorem~\ref{the:universalEncoder}), and let $\v{y}_0 \rightarrow \cdots \rightarrow \v{y}_t$ be the path taken by the encoder in $G$ ($\v{y}_0 = \psi(\v{x})$ and $\v{y}_t = f_k(\v{x})$). Notice that the only node amongst $\v{y}_0, \hdots, \v{y}_t$ in $S$ is $\v{y}_0$ since $d_{in}(\v{y}) = 0$ for any $\v{y} \in S$. By the design of Algorithm~\ref{alg:universalDecoder}, we find that the decoder traverses the transpose graph $G^T$ starting at $f_k(\v{x}) = \v{y}_t$ until a node in $S$ is reached. Since the maximal in-degree in $G$ is one, then we find that the maximal out-degree in $G^T$ is also one; thus, the path taken by the decoder is well-defined and coincides with the reverse of the path taken by the encoder. Further, since only $\v{y}_0$ is in $S$, then we find that the decoder terminates the while loop with $\v{y}_0$. Thus, the decoder returns $\psi^{-1}(\v{y}_0) = \psi^{-1}(\psi(\v{x})) = \v{x}$.
\label{the:universalDecoder}
\end{proof}
\end{theorem}

We now extend the analysis of Theorem~\ref{the:universalEncoder} and Theorem~\ref{the:universalDecoder} to demonstrate that the average time complexity of the encoder and decoder is $O(\abs{\Sigma}^r \cdot f(n))$. 

\begin{lemma}
The average number of iterations of the while loop in Algorithm~\ref{alg:universalEncoder} is at most $\abs{\Sigma}^r$.
\begin{proof}
Let $t(\v{x})$ be the number of iterations performed in $f_k(\v{x})$. As shown in Theorem~\ref{the:universalEncoder}, an execution of Algorithm~\ref{alg:universalEncoder} is equivalent to a path on $G$ of length $t(\v{x})$. Since $d_{in}(\v{y}) \leq 1$ for all $\v{y} \in \Sigma^n$ and since $\psi$ is injective, then we find that any two paths in $G$ taken by two different inputs are disjoint. As all of the paths are in $\Sigma^n$, then we find that the sum of the path lengths satisfies:
\begin{equation*}
    \sum_{\v{x} \in \Sigma^k} t(\v{x}) \leq \abs{\Sigma}^n \implies \frac{1}{\abs{\Sigma}^k} \cdot \sum_{\v{x} \in \Sigma^k} t(\v{x}) \leq \abs{\Sigma}^{n-k} = \abs{\Sigma}^r.
\end{equation*}
\end{proof}
\label{lemma:O1}
\end{lemma}

\begin{corollary}
The encoder and decoder possess $O(\abs{\Sigma}^r \cdot T(n))$ average time for $T(n)$ the maximal time complexity amongst $\varphi, \varphi^{-1}, \v{1}_S, \v{1}_{\mathcal{C}(n)}$.
\end{corollary}

\subsection{General Extensions}
\label{sec:universal:extensions}

This section presents two useful extensions of the universal iterative algorithms that will be utilized in the next sections. We begin with a construction that simplifies the criteria of Construction~\ref{const:universal} for single-symbol redundancy,

\begin{construction}
Let $\mathcal{C}(n) \subseteq \Sigma^{n}$ be a given constraint such that $\abs{\mathcal{C}(n)} \geq \abs{\Sigma}^{n-1}$. Given an injective function $\xi: \overline{\mathcal{C}(n)} \rightarrow \Sigma^{n-1}$ and an indicator function $\v{1}_{\mathcal{C}(n)} : \Sigma^{n} \rightarrow \{0,1\}$, there exists an efficient channel construction with one redundancy symbol and $O(T(n))$ average time complexity for $T(n)$ the maximal time complexity amongst $\xi, \xi^{-1}, \v{1}_{\mathcal{C}(n)}$.
\begin{proof}
We will demonstrate the desired construction as a special case of Construction~\ref{const:universal} with $r=1$ and $k=n-1$ by selecting a specific partition $S$ and embedding $\psi$. We select the following,
\begin{enumerate}
    \item Subset $S = \set{\v{x}1}{\v{x} \in \Sigma^{n-1}}$, satisfying the condition $\abs{\mathcal{C}(n)} \geq \abs{S} \geq \abs{\Sigma}^{n-1}$ as $\abs{\mathcal{C}(n)} \geq \abs{\Sigma}^{n-1}$ and $\abs{S} = \abs{\Sigma}^{n-1}$.
    \item Embedding $\psi : \Sigma^{n-1} \rightarrow S$ such that $\psi(\v{x}) = \v{x}1$, and $\v{1}_S(\v{x}) = x_n$ for $\v{x} = (x_1, x_2, \hdots, x_n)$. 
    \item Injective function $\varphi : \overline{\mathcal{C}}(n) \rightarrow \overline{S}$, such that $\varphi(\v{x}) = \xi(\v{x})0$ (the injective property follows from $\xi$ being injective), and using the given function $\v{1}_{\mathcal{C}(n)} : \Sigma^n \rightarrow \{0,1\}$.
\end{enumerate}
Therefore, according to the correctness of Construction~\ref{const:universal}, we find that there exists a channel construction for $\mathcal{C}(n)$ with $r=1$ and $O(\abs{\Sigma} \cdot T(n)) = O(T(n))$ average time complexity.
\end{proof}
\label{const:shrink}
\end{construction}

We continue by demonstrating a construction for a channel that is derived from the intersection of several channels. This is interesting for problems where the transmitted messages need to simultaneously satisfy several different constraints (e.g., a common combination in DNA storage is to be both almost-balanced and also avoid run lengths~\cite{erlich2017dna}). Naive attempts at combining constructions derived from each constraint are not possible since the constraints are inter-dependent and thus it is not clear that the algorithm will converge. Conversely, following the universal framework proposed in this paper, we find that a construction for channel intersections can also be derived with significant ease,

\begin{construction}[Channel Intersection]
Let $\mathcal{C}_1(n), \mathcal{C}_2(n), \hdots, \mathcal{C}_m(n) \subseteq \Sigma^{n}$ be given constraints such that $\forall i, \abs{\overline{\mathcal{C}_i(n)}} \leq \abs{\Sigma}^{n-1-\lceil \log_{\abs{\Sigma}}(m) \rceil}$. Given injective functions $\forall i, \xi_i: \overline{\mathcal{C}_i(n)} \rightarrow \Sigma^{n-\lceil \log_{\abs{\Sigma}}(m) \rceil - 1}$ and indicator functions $\forall i, \v{1}_{\mathcal{C}_i(n)} : \Sigma^{n} \rightarrow \{0,1\}$, there exists an efficient channel construction for $\mathcal{C}(n) = \bigcap_i \mathcal{C}_i(n)$ with one redundancy symbol and $O(m\cdot T(n))$ average time complexity for $T(n)$ the maximal time complexity amongst $\xi_1, \hdots, \xi_m, \xi_1^{-1}, \hdots, \xi_m^{-1}, \v{1}_{\mathcal{C}_1(n)}, \hdots, \v{1}_{\mathcal{C}_m(n)}$.
\begin{proof}
We will demonstrate the desired construction as a special case of Construction~\ref{const:shrink}:
\begin{enumerate}
    \item We verify that $\abs{\mathcal{C}(n)} \geq \Sigma^{n-1}$ as follows,
    \begin{equation*}
    \abs{\mathcal{C}(n)} = \abs{\bigcap_i \mathcal{C}_i(n)} \geq \abs{\Sigma}^n - \sum_i \abs{\overline{\mathcal{C}_i(n)}} \geq \abs{\Sigma}^n - m \cdot \abs{\Sigma}^{n - 1 - \lceil \log(m) \rceil} \geq \abs{\Sigma}^{n-1}
    \end{equation*}
    
    \item We propose the following injective function $\xi : \overline{\mathcal{C}(n)} \rightarrow \Sigma^{n-1}$, where $\xi(\v{x})$ is computed as follows: let $i$ be the minimal index such that $\v{x} \in \overline{\mathcal{C}_i(n)}$ (exists since $\v{x} \in \overline{\mathcal{C}(n)}$), then:
    \begin{equation*}
    \xi(\v{x}) = \xi_i(\v{x})\v{p}_i
    \end{equation*}
    where $\v{p}_i$ is the encoding of $i$ using $\lceil \log(m) \rceil$ symbols. 
    
    \item Indicator function $\v{1}_{\mathcal{C}(n)} : \Sigma^n \rightarrow \{0,1\}$ is computed as $\v{1}_{\mathcal{C}(n)}(\v{x}) = \bigwedge_i \v{1}_{\mathcal{C}_i(n)}(\v{x})$.
\end{enumerate}
Therefore, according to the correctness of Construction~\ref{const:shrink}, we find that there exists a channel construction for $\mathcal{C}(n)$ with $r=1$ and $O(m \cdot T(n))$ average time complexity for $T(n)$ the maximal time complexity amongst $\xi_1, \hdots, \xi_m, \xi_1^{-1}, \hdots, \xi_m^{-1}, \v{1}_{\mathcal{C}_1(n)}, \hdots, \v{1}_{\mathcal{C}_m(n)}$.
\end{proof}
\label{const:intersect}
\end{construction}

In Sections~\ref{sec:local} and~\ref{sec:global}, we primarily build upon Construction~\ref{const:shrink} to demonstrate the effectiveness of the universal framework for a variety of constraints. While it would be impractical to present all combinations of constraints for Construction~\ref{const:intersect}, we demonstrate a combination of constraints that address a previously-unsolved channel constraint important for DNA storage as an example. 

\section{Local Channel Constraints}
\label{sec:local}

This section explores a special case involving \emph{local} channel constraints that apply to \emph{substrings} of the transmitted messages. That is, $\mathcal{C}$ is defined according to a parametric set of substrings that may not appear in the transmitted words. We begin this section by defining a local channel constraint, continue by applying the universal construction to local channel constraints, and then conclude by evaluating the effectiveness of the construction on a variety of practical constraints.

\subsection{Definitions and Preliminaries}

We begin by defining the substring-avoiding (SA) constrained channel as follows,

\begin{definition}[SA]
The $\v{w}(n)$-substring-avoiding constrained channel is a parametric constrained channel $\mathcal{C}$ such that $\mathcal{C}(n)$ includes all words of length $n$ that do not contain $\v{w}(n)$ as a substring.
\end{definition}

We extend to multiple-substring-avoiding (MSA) vectors as follows,

\begin{definition}[MSA]
The $W(n)$-multiple-substring-avoiding constrained channel is a parametric constrained channel $\mathcal{C}$ such that $\mathcal{C}(n)$ includes all words of length $n$ that do not contain any word $\v{w} \in W(n)$ as a substring. We assume for simplicity that all $\v{w} \in W(n)$ are of length $\ell = \ell(n)$.\footnote{Otherwise, the shorter words in $W(n)$ can be padded with all possible combinations until the maximal length.} 
\end{definition}

This section tackles the MSA constrained channel, proposing a universal construction for the channel and expressing various practical constrained channels as MSA channels.

\subsection{Proposed Construction}
\label{sec:local:construction}

We propose the following construction for all MSA codes,

\begin{construction}[MSA]
For a given MSA constraint $W(n)$, and given (1) an injective function $\chi : W(n) \rightarrow \Sigma^{\ell'}$ such that $\ell' = \ell - \lceil \log_{\abs{\Sigma}}(n) \rceil - 1$, and (2) an indicator function $\v{1}_{W(n)} : \Sigma^\ell \rightarrow \{0,1\}$, there exists an explicit parametric construction with one redundancy symbol and $O(n \cdot T(\ell))$ average time complexity for $T(\ell)$ the maximal time complexity amongst $\chi, \chi^{-1}, \v{1}_{W(n)}$.

\begin{proof}
We show that this construction is a special case of Construction~\ref{const:shrink}. We select $\xi : \overline{\mathcal{C}(n)} \rightarrow \Sigma^{n-1}$ as follows: let $i$ be the index of the first substring of length $\ell$ in $\v{x}$ that belongs to $W(n)$ (there must exist such a substring as $\v{x} \in \overline{\mathcal{C}(n)}$), then 
    \begin{equation*}
        \xi(\v{x}) = \v{x}_1^{i-1}\;\v{x}_{i+\ell}^n\;\v{p}_i\;\chi(\v{x}_i^{i+\ell-1}),
    \end{equation*} 
    where $\v{x}_j^k$ is the substring of $\v{x}$ from $j$ to $k$ (inclusive) and $\v{p}_i$ is the encoding of $i$ using $\lceil \log(n) \rceil$ symbols. Since $\chi : W(n) \rightarrow \Sigma^{\ell'}$, then $\abs{\xi(\v{x})} = n - \ell + \lceil \log(n) \rceil + \ell' = n - 1$. We find that $\xi$ is injective since the window is chosen in a deterministic fashion and $\chi$ is injective. Thus, $\xi^{-1}$ exists and can be computed by performing $\chi^{-1}$ on the last $\ell'$ symbols, decoding $\v{p}_i$ from the next $\lceil \log(n) \rceil$ symbols, and finally reinserting the window into the message. Further, we select $\v{1}_{\mathcal{C}(n)} : \Sigma^n \rightarrow \{0,1\}$ as a function that returns 1 if at least one of the $n - \ell + 1$ windows in the message belongs to $W(n)$ (evaluated using $\v{1}_{W(n)}$). We find that the average time complexity is $O(n \cdot T(\ell))$ for encoding and $O(T(\ell) + n)$ for decoding since $\xi,\v{1}_{\mathcal{C}(n)}$ can be performed in $O(n \cdot T(\ell))$ time and $\xi^{-1}$ can be performed in $O(T(\ell) + n)$ time, respectively.
\end{proof}
\label{const:local}
\end{construction}

We notice that this construction guarantees a code with a single redundancy symbol for any $W(n)$ satisfying
\begin{equation}
\abs{W(n)} \leq \abs{\Sigma}^{\ell - \log(n) - 1} = \frac{1}{\abs{\Sigma} \cdot n}\cdot \abs{\Sigma}^\ell.
\end{equation}
Further, if an efficient algorithm for computing $\chi$ is provided, then the code also possesses efficient encoding and decoding. Otherwise, notice that an algorithm with $T(\ell) = O(\ell \cdot \log(\abs{W(n)}))$ time and $O(\ell \cdot \abs{W(n)})$ space always exists (binary search on $W(n)$).

\subsection{Applications}
\label{sec:local:applications}

We review in this section a wide variety of local channel constraints and demonstrate the effectiveness of the proposed universal constraint. We compare the proposed construction to the previous state-of-the-art algorithms in terms of the redundancy $r$ and the minimal $\ell$ supported (supporting smaller $\ell$ is equivalent to a more strict constraint and is thus desired). Further, we tackle a previously-unsolved problem of minimal palindromes that is also utilized in Section~\ref{sec:global} to solve an improved reverse-complement constraint that is likewise previously unsolved. 

\subsubsection{Minimal Hamming Weight}
\label{sec:local:applications:minimalHamming}

The minimal Hamming weight (MW) constraint requires that all $\ell$-substrings of the transmitted message possess some minimal Hamming weight $p(n)$, for $\Sigma=\{0,1\}$.\footnote{The case of $\abs{\Sigma} > 2$ follows in a straightforward manner.} The state-of-the-art previous work~\cite{levy2018mutually} is based on an iterative approach tailored to guarantee monotonic progression, requiring $p(n)$ bits of redundancy and supporting minimal $\ell$ of $\lceil \log(n) \rceil + (p(n) - 1) \cdot \lceil \log(\ell + 2) \rceil + 2$. Their algorithm begins by padding $p(n)$ ones to the end of the message, and then iteratively replaces windows with Hamming weight less than $p(n)$ with their alternate representations while additionally padding $10$ to guarantee monotonic progression in terms of the windows satisfied. 

We now apply the proposed construction to this channel. We first observe that the channel is an MSA constraint defined according to,
\begin{equation}
W_{\ell,p-MW}(n) = \set{\v{w} \in \Sigma^\ell}{w_H(\v{w}) < p(n)}.
\end{equation}
We design $\chi : W(n) \rightarrow \Sigma^{\ell'}$ as follows: for $\v{w} \in W(n)$, we define $\chi(\v{w}) = \v{p}_{i_1}\v{p}_{i_2}\cdots \v{p}_{i_{p(n)-1}}$ for $i_1, \hdots, i_{p(n)-1}$ the indices of the non-zero symbols\footnote{If $w_H(\v{w}) < p(n)-1$, then we append a dummy index of $\ell$ to represent no entry.} in $\v{w}$ and $\v{p}_i$ the binary encoding of $i$ using $\lceil \log(\ell + 1) \rceil$ bits. We thus require that,
\begin{equation*}
\ell' = (p(n) - 1) \cdot \lceil \log(\ell + 1) \rceil \leq \ell - \lceil \log(n) \rceil - 1
\end{equation*}
\begin{equation*}
\iff \ell \geq \lceil \log(n) \rceil + (p(n) - 1) \cdot \lceil \log(\ell + 1) \rceil + 1.
\end{equation*}

Therefore, we find that the proposed construction \emph{both} improves the minimal $\ell$ from $\lceil \log(n) \rceil + (p(n) - 1) \cdot \lceil \log(\ell + 2) \rceil + 2$ to $\lceil \log(n) \rceil + (p(n) - 1) \cdot \lceil \log(\ell + 1) \rceil + 1$, and more importantly also reduces the redundancy from $p(n)$ symbols to only one redundancy symbol.

\subsubsection{Local Almost-Balanced Hamming Weight}
\label{sec:local:applications:balancedHamming}

We now shift to considering the local almost-balanced Hamming weight (LAB) channel constraint that requires that the Hamming weight of all $\ell$-substrings be within the interval $[p_1\ell, p_2\ell]$. The previous state-of-the-art~\cite{nguyen2020binary} similarly constructs an iterative approach that consists of three phases (initial phase, replacement phase, and extension phase) tailored for this specific channel to guarantee monotonic convergence. They first prove that there exists a code for $\ell \geq \ln(n)/c^2$ with a single redundancy symbol, where $c=\min \left\{\frac{1}{2}-p_1, p_2-\frac{1}{2}\right\}$, yet they only successfully demonstrate $\ell \geq \ln(n)/c^2 + 2$ due to additional symbols utilized to guarantee the convergence. Conversely, our proposed construction successfully attains $\ell \geq \ln(n)/c^2$ with a single redundancy symbol.

We first express the channel constraint as an MSA constraint according to,
\begin{equation}
W_{\ell,p_1,p_2-LAB}(n) = \set{\v{w} \in \Sigma^\ell}{w_H(\v{w}) \notin [p_1\ell, p_2\ell]}.
\end{equation}
For $\chi$, we utilize the same function $F$ assumed in the previous work~\cite{nguyen2020binary}. We find that the conditions for Construction~\ref{const:local} are satisfied when $\ell \geq \ln(n)/c^2$ (see \cite{nguyen2020binary}), thereby attaining the results shown to be theoretically attainable in~\cite{nguyen2020binary}. Note that the proposed algorithm does not require the existence of the additional function $G$ as defined in~\cite{nguyen2020binary}, which is essential in the extension phase of the algorithm of~\cite{nguyen2020binary}. 

Overall, we demonstrate slightly improved $\ell$ with identical redundancy, while also assuming the existence of one less function since we do not need to perform extension. 

\subsubsection{Minimal Periodicity}
\label{sec:local:applications:periodicity}

We begin by explicitly defining periodicity of strings,

\begin{definition} For $\v{w} \in \Sigma^\ell$, $1 \leq p \leq \ell-1$ is called a \emph{period} of $\v{w}$ if for all $1 \leq i \leq \ell-p, w_{i} = w_{i+p}$.
\label{def:period}
\end{definition}

The minimal periodicity (MP) channel constraint requires that all $\ell$-windows possess a minimal periodicity of at least $p(n)$. Notice that this channel is an extension of the well-known run-length-limited (RLL) constraint~\cite{levy2018mutually} which constrains long sequences comprised of the repetition of the same symbol ($p(n)=2$ is similar to RLL). A recent work~\cite{ConstrainedPeriodicity} demonstrated near-optimal results based on the approach that is generalized in this work, attaining one redundancy bit for $\ell$ of at least $\lceil \log(n) \rceil + p + 1$, significantly improving the results of the previous state-of-the-art~\cite{MultipleHeadRacetrack} in both redundancy and minimal $\ell$. The \emph{generalized} construction proposed in this paper coincides with these state-of-the-art results for MSA constraint defined according to,
\begin{equation}
W_{\ell,p-MP}(n) = \set{\v{w} \in \Sigma^\ell}{\exists p' < p(n)\text{ such that $\v{w}$ has minimal period $p'$}},
\end{equation}
for $\chi(\v{w})$ chosen as the concatenation of the first $p'$ bits of $\v{w}$, a single one symbol, and $p(n)-p'-1$ zero symbols (used to encode both $p'$ and the first $p'$ symbols from which the window can be retrieved~\cite{ConstrainedPeriodicity}):
\begin{equation*}
\chi(\v{w}) = \v{w}_1^{p'}10^{p(n)-p'-1}.
\end{equation*}

Therefore, we find that overall the proposed algorithm successfully generalizes~\cite{ConstrainedPeriodicity} to beyond only periodicity while still coinciding with the state-of-the-art results for periodicity.

\subsubsection{Minimal Palindromes}
\label{sec:local:applications:palindromes}

The exact no palindromes (ENP) constraint restricts that there are no palindromes of length exact $\ell$ in the message of length $n$. The constraint has applications in DNA storage since palindromes have a negative impact on the DNA sequencing pipelines~\cite{warris2018correcting}. Nonetheless, to the best of our knowledge, the constraint is previously unsolved -- potentially due to the complexity of the extension step in the sequence-replacement technique. We express the constraint as an MSA constraint as follows,
\begin{equation}
W_{\ell-ENP}(n) = \set{\v{w} \in \Sigma^\ell}{\v{w}^R = \v{w}},
\end{equation}
where $(w_1, \hdots, w_\ell)^R = (w_\ell, \hdots, w_1)$. For $\chi : W(n) \rightarrow \Sigma^{\ell'}$, we utilize $\chi(\v{w}) = \v{w}_1^{\lceil \ell/2 \rceil}$ since the remainder of the string $\v{w}_{\lceil \ell/2 \rceil + 1}^n$ can be derived from $\v{w}_1^{\lceil \ell/2 \rceil}$ as $\v{w}^R = \v{w}$. We thus require that,
\begin{equation}
\ell' = \lceil \ell/2 \rceil \leq \ell - \lceil \log(n) \rceil - 1 \iff \lfloor \ell / 2 \rfloor = \ell - \lceil \ell/2\rceil \geq \lceil \log(n) \rceil + 1
\end{equation}
Therefore, we find that there exists an efficient construction with a single redundancy symbol for $\lfloor \ell / 2 \rfloor \geq \lceil \log(n) \rceil + 1$. Notice that this also addresses all palindromes of length $\ell + 2k$ for $k \in \mathbb{N}$ since any palindrome of length $\ell + 2k$ contains a palindrome of length $\ell$. To solve the general minimal palindromes (MPL) constraint that restricts any palindromes of length $\geq \ell$ in the message of length $n$, we find that Construction~\ref{const:intersect} can be utilized to develop a construction for MPL with a single redundancy symbol and $\ell = 2 \cdot \lceil \log(n) \rceil + 4$ (intersecting ENP constructions for even and odd $\ell$). 

\section{Global Channel Constraints}
\label{sec:global}

In this section we consider two \emph{global} channel constraints operating on pairs of subsequences in the overall transmitted message. We begin with the repeat-free constraint~\cite{elishco2021repeat} which eliminates repetitions of the same subsequence in two or more locations in the transmitted message, and continue with the secondary structure constraint (DNA reverse complement) that generalizes the repeat-free constraint. We further improve the construction for the secondary-structure constraint to also address overlapping windows, thereby solving a previously-unsolved problem important for DNA-storage applications~\cite{warris2018correcting, nguyen2023design}. Lastly, we tackle the open problem of the global almost-balanced Hamming weight channel~\cite{nguyen2020binary} for weight margin $\pm\sqrt{n}$ and demonstrate an efficient algorithm with a single redundancy symbol.

\subsection{Repeat-Free}
\label{sec:global:repeatFree}

The repeat-free constraint requires that any $\ell$-substring of the transmitted message appear exactly once in the message (i.e., there are no repeated substrings of length $\ell$). In \cite{elishco2021repeat}, the authors thoroughly studied the repeat-free constraint and proposed a construction that utilizes \emph{two redundancy symbols} to support $\ell \geq 2\cdot \lceil \log(n) \rceil + 2$. The construction is based on an iterative approach with elimination and expansion phases (similar to the other previous works), yet they further constrain the algorithm to also require that there are no $\log(n)+1$ zero runs since this is necessary for the correctness of the expansion phase. Conversely, we demonstrate $\ell \geq 2 \cdot \lceil \log(n) \rceil + 1$ with only a \emph{single redundancy symbol} by applying Construction~\ref{const:shrink}.

We first explicitly define the repeat-free (RF) constraint as follows,
\begin{equation}
\mathcal{C}_{\ell-RF}(n) = \set{\v{x} \in \Sigma^n}{\nexists 1 \leq i < j \leq n - \ell(n) + 1 \;:\; \v{x}_i^{i + \ell(n) - 1} = \v{x}_j^{j+\ell(n)-1}},
\end{equation}
for $\ell = \ell(n)$. We apply Construction~\ref{const:shrink} to this problem to find,

\begin{construction}[Repeat-Free]
There exists an efficient construction with a single redundancy symbol for $\mathcal{C}_{\ell-RF}(n)$ when $\ell = 2 \cdot \lceil\log(n)\rceil + 1$, with $O(n \log^2(n))$ average time complexity.
\begin{proof}
We will demonstrate the desired construction as a special case of Construction~\ref{const:shrink}. We select $\xi: \overline{\mathcal{C}_{\ell-RF}(n)} \rightarrow \Sigma^{n-1}$ as follows: let $1 \leq i < j \leq n - \ell + 1$ be the minimal\footnote{The order is defined by a primary order on $i$ and a secondary order on $j$.} indices such that $\v{x}_i^{i + \ell-1} = \v{x}_j^{j+\ell-1}$ (must exist as $\v{x} \in \overline{\mathcal{C}_{\ell-RF}(n)}$), then we define $\xi(\v{x})$ as $\v{x}$ after removing the window at $j$ and appending encodings of $i$ and $j$,
\begin{equation*}
\xi(\v{x}) = \v{x}_1^{j-1}\v{x}_{j+\ell}^{n}\v{p}_i\v{p}_j
\end{equation*}
where $\v{p}_m$ is the encoding of $m$ using $\lceil \log(n) \rceil$ symbols. We find that $\abs{\xi(\v{x})} = n - \ell + 2 \cdot \lceil \log(n) \rceil = n - 1$, as desired. We show that $\xi$ is injective by proposing $\xi^{-1}$ as follows: $\xi^{-1}(\v{y})$ first removes and decodes the last $2 \cdot \lceil \log(n) \rceil$ symbols of $\v{y}$ to find $i$ and $j$, and then:
\begin{itemize}
    \item \emph{If $j \geq i + \ell$ (no overlap)}, then $\xi^{-1}$ copies the window at index $i$ of $\v{y}$ and inserts it into position $j$. In this case, we find that the proposed $\xi^{-1}$ successfully inverts $\xi$ since the windows at $i$ and $j$ were identical in $\v{x}$. 
    \item \emph{If $j < i + \ell$ (overlap)}, then we find that $\v{w} = \v{x}_i^{i + \ell - 1}$ was periodic with period $j-i$ since $\v{w}_{(j - i)+1}^{\ell} = \v{w}_1^{\ell - (j-i)}$. Therefore, $\xi^{-1}$ reconstructs $\v{w}$ from $\v{y}_i^{i+(j-i)-1} = \v{x}_i^{i+(j-i)-1} = \v{w}_1^{j-i}$ (using the fact that $\v{w}$ is $j-i$ periodic) and then inserts $\v{w}$ at index $j$ into $\v{y}$. The proposed $\xi^{-1}$ successfully inverts $\xi$ since it recovers the window at $j$ and then reinserts it. 
\end{itemize}

We select $\v{1}_{\mathcal{C}_{\ell-RF}(n)}$ as an algorithm that checks if there exist two repeating $\ell$-substrings in $O(n \log(n) \cdot \ell) = O(n\log^2(n))$ time complexity (e.g., a tree-based set). Based on Construction~\ref{const:shrink}, we find that the overall average time complexity is $O(n \log^2(n))$ for both encoding and decoding.
\end{proof}
\label{const:rf}
\end{construction}

We notice that Construction~\ref{const:rf} can be generalized with the same parameters to the symbolwise-repeat-free (SRF) constraint, defined as follows:
\begin{equation}
\mathcal{C}_{\ell,\alpha-SRF}(n) = \set{\v{x} \in \Sigma^n}{\nexists 1 \leq i < j \leq n - \ell + 1 \;:\; \alpha(\v{x}_i^{i + \ell - 1}) = \v{x}_j^{j+\ell - 1}}
\end{equation}
for $\alpha : \Sigma^\ell \rightarrow \Sigma^\ell$ a symbolwise function (i.e., $\alpha(x_1, \hdots, x_\ell) = (\beta_1(x_1), \hdots, \beta_\ell(x_\ell))$ for $\forall i, \beta_i : \Sigma \rightarrow \Sigma$).

\subsection{Secondary Structure (DNA Reverse Complement)}
\label{sec:global:secondary}

The secondary structure channel constraint arises from the field of DNA storage~\cite{milenkovic2006design}, whereby data is encoded in synthesized DNA strands using the alphabet $\Sigma = \{A, T, C, G\}$. To prevent the DNA strand from folding onto itself, it is necessary that there are no two windows in the strand that are the reverse-complement (RC)\footnote{Note that any $c : \Sigma \rightarrow \Sigma$ such that $c = c^{-1}$ is supported.} of each-other:
\begin{equation}
RC(x_1, \hdots, x_n) = (c(x_n), \hdots, c(x_1)), \quad\text{where}\quad c(\alpha) = \begin{cases} A & \alpha = T \\ T & \alpha = A \\ C & \alpha = G \\ G & \alpha = C \end{cases}
\end{equation}

The state-of-the-art construction~\cite{nguyen2023design} solves a relaxed problem with only non-overlapping\footnote{Overlapping windows are more difficult to address due to the reverse complement no longer implying periodicity.} windows, and is based on an iterative construction with elimination and expansion phases. They perform expansion with padding of the form $(AC)^{({n-N_0})/{2}}$, which requires them to further constrain the message to not include any patterns of the form $(\alpha_1\alpha_2)^{({3\log n + 2})/{2}}$ for $\alpha_1, \alpha_2 \in \Sigma$. Overall, for $\ell \geq 3 \cdot \lceil \log(n) \rceil + 4$, they propose a construction with a single redundancy symbol. We demonstrate that the proposed construction can match the single-symbol redundancy for this relaxed constraint while reducing the minimal $\ell$ from $3 \cdot \lceil \log(n) \rceil + 4$ to $2 \cdot \lceil \log(n) \rceil + 1$. 

We begin by explicitly defining the relaxed secondary structure (RSS) constraint,
\begin{equation}
\mathcal{C}_{\ell-RSS}(n) = \set{\v{x} \in \Sigma^n}{\nexists 1 \leq i < n - \ell + 1, i + \ell \leq j \leq n - \ell + 1 \;:\; RC(\v{x}_i^{i + \ell - 1}) = \v{x}_j^{j+\ell-1}}.
\end{equation}
Similar to Construction~\ref{const:rf}, we find that applying Construction~\ref{const:shrink} to the problem yields the following results:

\begin{construction}[Relaxed Secondary Structure]
There exists an efficient construction with a single redundancy symbol for $\mathcal{C}_{\ell-RSS}(n)$ when $\ell = 2 \cdot \lceil\log(n)\rceil + 1$, with $O(n \log^2(n))$ average time complexity.
\begin{proof}
Similar to the proof of Construction~\ref{const:rf}, without needing to address overlapping windows and using the RC function instead of equality.
\end{proof}
\label{const:rss}
\end{construction}

In order to address also overlapping substrings, we notice that overlapping substrings constitute a \emph{local} constraint since they are overlapping. Specifically, we find that an overlap of an $\ell$-substring with its reverse complement guarantees the existence of a palindrome of length $2 \cdot \lfloor\ell/2\rfloor + 2$ in the message. Therefore, by utilizing Construction~\ref{const:intersect} and Section~\ref{sec:local:applications:palindromes} we find the following construction,

\begin{equation}
\mathcal{C}_{\ell-SS}(n) = \set{\v{x} \in \Sigma^n}{\nexists 1 \leq i < j \leq n - \ell + 1 \;:\; RC(\v{x}_i^{i + \ell - 1}) = \v{x}_j^{j+\ell - 1}}.
\end{equation}
\begin{construction}[Secondary Structure]
There exists an efficient construction with a single redundancy symbol for $\mathcal{C}_{\ell-SS}(n)$ when $\ell = 2 \cdot \lceil\log(n)\rceil + 2$, with $O(n \log^2(n))$ average time complexity.
\label{const:ss}
\end{construction}

\subsection{Almost-Balanced Hamming Weight}
\label{sec:global:almostBalanced}

We consider here the global almost-balanced constraint which generalizes the well-known balanced Knuth codes~\cite{Knuth} by requiring that the entire message possess a Hamming weight of \emph{approximately} $n/2$. This construction has wide implications to applications such as DNA storage, where balanced GC content is necessary~\cite{blawat2016forward}. It follows from a combinatorical analysis that there exists a single-redundancy-symbol construction for the channel consisting of all words of length $n$ with Hamming weight $[n/2-\sqrt{n}/2, n/2+\sqrt{n}/2]$ for large enough $n$\footnote{This can be seen from the fact that the binomial distribution approaches the normal distribution as $n \rightarrow \infty$, and that at least half of the space is thus contained in $[\mu-\sigma, \mu+\sigma] = [n/2-\sqrt{n}/2, n/2+\sqrt{n}/2]$.}. Nonetheless, the problem has remained unsolved as the state-of-the-art construction~\cite{nguyen2020binary} only tackles a linear-almost-balanced version of $[np_1, np_2]$ for $p_1 < 1/2 < p_2$. We below demonstrate an efficient construction with a single redundancy symbol for $[n/2-\sqrt{n}, n/2+\sqrt{n}]$ based on Construction~\ref{const:intersect} and the arithmetic coding~\cite{rissanen1979arithmetic} technique. We first explicitly define the constraint as follows,

\begin{equation}
\mathcal{C}_{AB}(n) = \set{\v{x} \in \Sigma^n}{w_H(\v{x}) \in \left[\frac{n}{2}-\sqrt{n},\frac{n}{2}+\sqrt{n}\right]}.
\end{equation}

We also define the two following helper constraints for the overall construction,

\begin{equation}
\mathcal{C}_{AB-L}(n) = \set{\v{x} \in \Sigma^n}{w_H(\v{x})  \leq \frac{n}{2}+\sqrt{n}},
\end{equation}
\begin{equation}
\mathcal{C}_{AB-H}(n) = \set{\v{x} \in \Sigma^n}{w_H(\v{x}) \geq \frac{n}{2}-\sqrt{n}}.
\end{equation}

Notice that $\mathcal{C}_{AB} = \mathcal{C}_{AB-L}(n) \cap \mathcal{C}_{AB-H}(n)$. We now propose the overall construction as follows,

\begin{construction}[Almost-Balanced]
There exists an efficient construction with a single redundancy symbol for $\mathcal{C}_{AB}(n)$ when $n > 4$.
\begin{proof}
We demonstrate this construction as a special case of Construction~\ref{const:intersect} utilizing $\mathcal{C}_{AB} = \mathcal{C}_{AB-L}(n) \cap \mathcal{C}_{AB-H}(n)$. Without loss of generality, we design the injective function $\xi_1 : \overline{\mathcal{C}_{AB-H}} \rightarrow \Sigma^{n-2}$ as follows: given a word $\v{x} \in \overline{\mathcal{C}_{AB-H}} \iff w_H(\v{x}) < n/2 - \sqrt{n}$, we utilize \emph{arithmetic coding}~\cite{rissanen1979arithmetic}\footnote{Arithmetic coding is an iterative approach that is based on splitting the real interval $[0,1)$ at each iteration into $[0,p)$ and $[p,1)$, where the selection of the interval is based on the current input bit. By showing that the intervals of select words are at least of length $\ell > 0$, it is possible to enumerate those words with $\lceil \log(1/\ell) \rceil$ symbols.} with $p=1/2+1/\sqrt{n}$ to generates intervals of length,
\begin{equation*}
\left(\frac{1}{2} - \frac{1}{\sqrt{n}}\right)^{w_H(\v{x})} \cdot \left(\frac{1}{2} + \frac{1}{\sqrt{n}}\right)^{n - w_H(\v{x})} \geq 1/\abs{\Sigma}^{n-2}
\end{equation*}
for  the worst case of $w_H(\v{x}) = n/2 - \sqrt{n} - 1$ as long as $n>4$. Therefore, according to the correctness of arithmetic coding, we find that we have successfully defined $\xi_1$ as an injective function from $\overline{\mathcal{C}_{AB-H}}$ to $\Sigma^{n-2}$. 

Thus, we find that there exists the desired $\xi_1 : \overline{\mathcal{C}_{AB-H}} \rightarrow \Sigma^{n-2}$ and $\xi_2 : \overline{\mathcal{C}_{AB-L}} \rightarrow \Sigma^{n-2}$. Therefore, we find according to Construction~\ref{const:intersect} that there exists an efficient construction for $\mathcal{C}_{AB} = \mathcal{C}_{AB-L}(n) \cap \mathcal{C}_{AB-H}(n)$ with a single bit of redundancy.

\end{proof}
\label{const:ab}
\end{construction}

\section{Conclusion}
\label{sec:conclusion}

Parametric constrained coding generalizes the well-known field of fixed constrained coding to address constraints that grow with the message length and portray some general property (e.g., avoid windows of $\log(n)$ consecutive zeros). While there is a universal solution to any \emph{fixed} constrained coding task, the lack of such a solution for \emph{parametric} constraints has led to custom algorithms being designed for each constraint. Specifically, most approaches are based on iteratively replacing invalid substrings in the transmitted message, yet then they require complex custom transition functions that enable the monotonic progression which guarantees convergence. In this work, we propose a novel universal framework for all parametric constrained coding problems that is able to generalize to all constraints due to a surprising proof that demonstrates that there is no need for monotonic progression to guarantee efficient convergence. We apply this universal constraint to both local and global parametric constraints, and demonstrate significant improvements to the state-of-the-art constructions for a wide variety of problems which significantly simplifies the algorithms. Further, we demonstrate the first construction for several previously-unsolved constraints. Therefore, this universal framework may also be readily applied in the future to tackle additional previously-unsolved constrained coding problems with significant ease.

\bibliographystyle{alpha}
\bibliography{refs}
\end{document}